\documentclass[twocolumn]{aa}
\usepackage{graphicx}
\usepackage{amssymb}
\begin{document}

   \title{The VIMOS-VLT Deep Survey}
\subtitle{Luminosity dependence of clustering at  $z \simeq 1$\thanks{based on data obtained with the European Southern Observatory Very Large
    Telescope, Paranal, Chile, program 070.A-9007(A), and on data
    obtained at the Canada-France-Hawaii Telescope, operated by the
    CNRS of France, CNRC in Canada and the University of Hawaii. }
}

\author{A. Pollo \inst{1,2,15}
\and L. Guzzo \inst{2}
\and O. Le F\`evre \inst{1}
\and B. Meneux \inst{1}
\and A. Cappi    \inst{3}
\and P. Franzetti \inst{4}
\and A. Iovino \inst{2}
\and H.J. McCracken \inst{5,6}
\and C. Marinoni \inst{7,1}
\and G. Zamorani \inst{3}
\and D. Bottini \inst{4}
\and B. Garilli \inst{4}
\and V. Le Brun \inst{1}
\and D. Maccagni \inst{4}
\and J.P. Picat \inst{8}
\and R. Scaramella \inst{9}
\and M. Scodeggio \inst{4}
\and L. Tresse \inst{1}
\and G. Vettolani \inst{10}
\and A. Zanichelli \inst{10}
\and C. Adami \inst{1}
\and S. Arnouts \inst{1}
\and S. Bardelli  \inst{3}
\and M. Bolzonella  \inst{14} 
\and S. Charlot \inst{5}
\and P. Ciliegi    \inst{3}  
\and T. Contini \inst{8}
\and S. Foucaud \inst{12}
\and I. Gavignaud \inst{8,13}
\and O. Ilbert \inst{14}
\and B. Marano     \inst{14}  
\and A. Mazure \inst{1}
\and R. Merighi   \inst{3} 
\and S. Paltani \inst{19,20}
\and R. Pell\`o \inst{8}
\and L. Pozzetti    \inst{3} 
\and M. Radovich \inst{17}
\and E. Zucca    \inst{3}
\and M. Bondi \inst{10}
\and A. Bongiorno \inst{14}
\and G. Busarello \inst{17}
\and O. Cucciati \inst{2,18}
\and L. Gregorini \inst{10}
\and F. Lamareille \inst{8}
\and G. Mathez \inst{8}
\and Y. Mellier \inst{5,6}
\and P. Merluzzi \inst{17}
\and V. Ripepi \inst{17}
\and D. Rizzo \inst{16}
}

\offprints{agnieszka.pollo@oamp.fr}

\institute{
Laboratoire d'Astropysique de Marseile, UMR 6110 CNRS-Universit\'e de
Provence,  BP8, 13376 Marseille Cedex 12, France
\and
INAF - Osservatorio Astronomico di Brera - Via Bianchi 46, I-23807,
Merate \& Via Brera 28, I-20121, Milano, Italy
\and
INAF - Osservatorio Astronomico di Bologna - Via Ranzani,1, I-40127, Bologna, Italy
\and
INAF - IASF Milano - via Bassini 15, I-20133, Milano, Italy
\and
Institut d'Astrophysique de Paris, UMR 7095, 98 bis Bvd Arago, 75014
Paris, France
\and
Observatoire de Paris, LERMA, 61 Avenue de l'Observatoire, 75014 Paris, 
France
\and
Centre de Physique Th\'eorique CNRS-Luminy and Universit\'e de Provence, UMR
6207,  F-13288 Marseille Cedex 9, France
\and
Laboratoire d'Astrophysique de l'Observatoire Midi-Pyr\'en\'ees (UMR 
5572) - 14, avenue E. Belin, F31400 Toulouse, France
\and
INAF - Osservatorio Astronomico di Roma - Via di Frascati 33,
I-00040, Monte Porzio Catone,
Italy
\and
INAF - IRA - Via Gobetti,101, I-40129, Bologna, Italy
\and
Max Planck Institut fur Astrophysik, 85741, Garching, Germany
\and
School of Physics \& Astronomy, University of Nottingham, University
Park, Nottingham, NG72RD, UK
\and
Universitat Postdam, Astrophysik, 14469 Postdam, Germany
\and
Universit\`a di Bologna, Dipartimento di Astronomia - Via Ranzani,1,
I-40127, Bologna, Italy
\and
Astronomical Observatory, Jagiellonian University, ul. Orla 171, 30-244 Krak\'ow, Poland
\and
Imperial College London, South Kensington Campus, London SW7 2AZ, UK
\and
INAF-Osservatorio Astronomico di Capodimonte - Via Moiariello 16, I-80131, Napoli, Italy
\and
Universit\'a di Milano-Bicocca, Dipartimento di Fisica -
Piazza delle Scienze, 3, I-20126 Milano, Italy
\and
Integral Science Data Centre, ch. d'\'Ecogia 16, CH-1290 Versoix, Switzerland
\and
Geneva Observatory, ch. des Maillettes 51, CH-1290 Sauverny, Switherland
}

   \date{Received ; accepted }

   \abstract{We investigate the dependence of galaxy clustering on the
   galaxy intrinsic luminosity at high redshift, using the 
data from the First Epoch VIMOS-VLT Deep Survey (VVDS).  
The size (6530 galaxies)
and depth ($I_{AB}<24$) of the survey allows us to measure the projected
two-point correlation function of galaxies, $w_p(r_p)$,
for a set of volume-limited samples
up to an effective redshift $\left<z\right>=0.9$
and median absolute magnitude $-19.6< M_B < -21.3$.
Fitting $w_p(r_p)$ with a single power-law model for the real-space
correlation function $\xi(r)=(r/r_0)^{-\gamma}$, we measure the
relationship of the correlation length $r_0$ and the slope $\gamma$
with the sample median luminosity for the first time at such high
redshift.  Values from our lower-redshift samples ($0.1<z<0.5$) are fully consistent with the
trend observed by larger local surveys. 
In our high redshift sample ($0.5<z<1.2$), we find that the clustering
strength suddenly rises around $M_B^*$, apparently with a sharper
inflection than at low redshifts.  Galaxies in the faintest
sample ($\left<M_B\right>=-19.6$) have a correlation length 
$r_0=2.7^{+0.3}_{-0.3}$ $h^{-1}$
Mpc, compared to $r_0=5.0^{+1.5}_{-1.6}$ $h^{-1}$ Mpc at $\left<M_B\right>=-21.3$.
The slope of the correlation
function is observed  
to correspondingly steepen significantly
from $\gamma=1.6^{+0.1}_{-0.1}$ 
to $\gamma=2.4^{+0.4}_{-0.2}$. 
This is not observed either by large local surveys or in our
lower-redshift samples and seems to imply a significant change in the way
luminous galaxies trace dark-matter halos at $z\sim 1$ with respect to
$z\sim 0$.
At our effective median redshift $z \simeq 0.9$ 
this corresponds to a strong difference of the relative bias,
from $b/b* < 0.7$ for galaxies with $L < L*$ to $b/b* \simeq 1.4$ for
galaxies with $L > L*$.   

   }

   \maketitle
%

\section{Introduction}

At the current epoch, luminous galaxies tend to be more clustered than
faint ones (\cite{davis}, \cite{hamilton}, \cite{iovino}, 
\cite{maurogordato}, \cite{benoist}, \cite{willmer98}, 
Guzzo et al., 2000, 
\cite{norberg}, \cite{norberg02}, \cite{zehavi}), with the
difference becoming remarkable above the characteristic luminosity
$L_*$ of the Schechter luminosity function.
This effect is in general agreement with predictions from
hierarchical models of galaxy formation (\cite{white}, \cite{valls}, 
Kauffmann et al. 1997,
\cite{benson}), in which bright galaxies are expected to occupy more 
massive dark matter haloes than fainter ones and these haloes are 
more strongly clustered than the overall distribution of dark matter
(\cite{kaiser84, mo-white, sheth-tormen}). 
If this is the case, the difference in clustering between faint and
bright galaxies should become even more evident at high redshifts,
where galaxy formation is supposed to be more confined to the highest
peaks of the density field. 

Understanding the relationship between galaxies and dark matter halos
is one of the most difficult challenges of the theory in
predicting the observed clustering of galaxies.  Over the last few
years, {\sl halo occupation models} have provided this connection in a
phenomenological way, allowing one, e.g., to explain the detailed shape of
the galaxy two-point correlation function (\cite{zehavi04}, but see
also \cite{guzzo91}).  In these models, a 
statistically motivated recipe
to describe galaxy formation 
determines the halo occupation
distribution (HOD), specifying the probability $P(N|M)$ that a halo of
virial mass $M$ contains $N$ galaxies of a given type, together with
any spatial and velocity biases of galaxies
(\cite{kauff}, \cite{benson}, \cite{berlind03}, \cite{kravtsov04}).
This term (known as the {\it one-halo component} 
of the correlation function) governs the behaviour 
of galaxy correlations on small ($<2$
$h^{-1}$ Mpc) scales, while at larger separations galaxy correlations
are dominated by the gravitational clustering of virialized dark
matter halos (the {\it two-halo component}), 
with essentially no dependence on the more complex
physics of the sub-dominant baryonic component.  Given cosmological
parameters and a specified HOD, therefore, one can calculate any galaxy
clustering statistic, on any scale (e.g., \cite{abazajian}), either by
populating the halos from an N-body simulation (e.g.,
\cite{jing98}, \cite{jing02}) or via analytical prescriptions (e.g.,
\cite{peacock00}, \cite{seljak00}, \cite{marinoni02}, \cite{cooray02}, \cite{vdb03}). On the other hand, as it has been shown (\cite{sheth2004}, \cite{gao2005}, \cite{harker2005}), there seems to exist a clear relationship
between halo formation properties and halo clustering properties, which 
indicates that current HOD models may describe galaxy 
clustering at best approximately. Thus, observations of the relative clustering
of galaxies with different intrinsic luminosities provide crucial
constraints on HOD models. 

The detailed luminosity dependence of
clustering so far has been difficult to establish 
because of the limited dynamic range in luminosity for even the
largest local galaxy redshift surveys (e.g. \cite{norberg}). It is even
more problematic to study this effect at redshifts significantly
different to zero.  High redshift samples have been too small
to allow sub-division into luminosity classes.  An
additional complication relates to evolution of the overall luminosity
function: galaxies become brighter on average going back in time, thus
comparison of high-redshift measurements to local values requires 
accurate knowledge of the evolution of the global luminosity function.

The VIMOS-VLT Deep Survey (VVDS) provides us with unique 
information to address these issues
in detail.  A first investigation of how the non-linear bias between
galaxy and matter evolves with
redshift for different luminosity classes has been presented in
Marinoni et al. (2005). An analysis of the evolution of 
clustering of galaxies has been presented in Le F\`evre 
et al. (2005a), and the evolution of the dependence 
of clustering on spectral types has been
discussed by Meneux et al. (2005). In this paper we use
the same VVDS first-epoch data to measure in more detail 
the dependence of galaxy clustering on luminosity at
$\left<z\right>\simeq 0.9$, and compare it to local values from 2dFGRS and SDSS.
We describe the VVDS catalog and the construction of volume
limited samples in Section 2. Section 3 presents the methods to
estimate and retrieve the best-fit parameters for the real-space
correlation function. We present our results on the projected correlation 
function in Section 4, while the comparison to existing
local surveys, together with a discussion of the results is given in Section 5.

Throughout this paper we use a Concordance Cosmology with $\Omega_m =
0.3$ and $\Omega_{\Lambda} = 0.7$. The Hubble constant is normally
parameterised via $h=H_0/100$ to ease comparison with previous works,
while a value $H_0 = 70$ km s$^{-1}$ Mpc$^{-1}$ is used when
computing absolute magnitudes. All correlation length values are
quoted in comoving coordinates. 

\section{The data}
   \begin{figure}
   \centering
   \includegraphics[width=9cm]{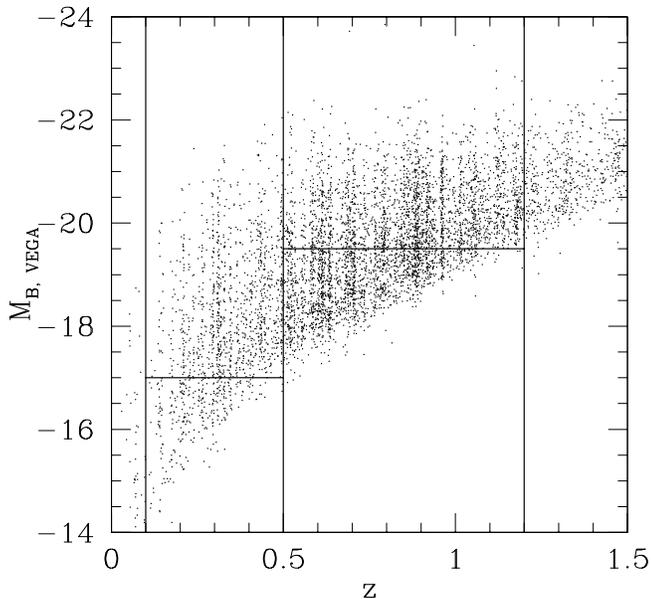}
      \caption{Distribution of magnitudes of VVDS galaxies with redshift. Solid vertical lines show the boundaries of our two redshift slices. Solid horizontal lines show from which magnitudes ($M_B \sim 19.5$) our sub-samples are volume limited.
}
         \label{magn}
   \end{figure}

\subsection{The VVDS}

The VIMOS-VLT Deep Survey (VVDS, \cite{LEF04}) studies
the evolution of galaxies and the large scale structure of the Universe to
$z \sim 2$ and higher. The VVDS spectroscopic survey is performed
with the VIMOS multi-object spectrograph at the
European Southern Observatory Very Large Telescope (\cite{LEF03}) and
complemented with multi-color BVRI imaging data obtained at the CFHT
telescope (\cite{hjmcc}, \cite{LEF04b}). The first-epoch VVDS data
consist of more than $11000$ spectra obtained in two VVDS-Deep fields,
covering $0.61$ square degrees.

For this analysis, we use the data from the F02 ``Deep'' field, which
is a purely magnitude limited survey to $I_{AB}=24$ currently covering
an area of $0.49$ square degrees.  Considering only galaxies with
secure ($>80\%$ confidence) redshifts, this sample includes $6530$
galaxies. Details about observations, data reduction, redshift measurement and
quality assessment can be found in Le F\`evre et al. (2005b).

\subsection{Luminosity-limited sub-samples}

To measure the dependence of clustering on galaxy luminosity,  we select
two redshift slices to isolate comparable intervals of cosmic time.
By choosing the intervals $z \in [0.1,0.5]$ and $z \in [0.5,1.2]$, we obtain two samples covering  approximately $3.5$ Gy of proper time, in the adopted cosmology.  The high-redshift slice is compared to the low-redshift sample 
and to local samples like the 2dF Galaxy Redshift Survey (2dFGRS) and
the Sloan Digital Sky Survey (SDSS). The low-redshift
sample is also directly compared to existing
local estimates, although the small volume and larger redshift limit
do not allow a one-to-one match.
Within each slice,  we build a series of volume-limited (where
statistically possible) or quasi-volume-limited sub-samples chosen 
to contain a statistically reasonable number of galaxies.   

Absolute magnitudes for VVDS galaxies have been estimated by computing
the appropriate K-correction via a SED fitting technique to the
observed multi-band photometry (see \cite{ilbert} and \cite{franzetti}
for details) and to ease comparison to previous work are reported here in
the VEGA system. 

Due to the apparent magnitude limit of the survey, as shown in Figure
\ref{magn}, within the high-redshift slice a true volume-limited
sample can be defined only for $M_B \lesssim
-19.5$. Conversely, in the low-redshift slice we can define
volume-limited sub-samples for $M_B \lesssim -17$. Because of the
smaller volume probed we do not have a large number of bright galaxies
in this latter sample.  Therefore, as seen from Figure \ref{magn},
we cannot expect statistically sound measurements for 
galaxies closer than $z \sim 0.5$ with $M_B \lesssim -19.5$.  
All details on the selected VVDS sub-samples are listed in Tables 1 and 2.
For each
sub-sample, the columns give its absolute magnitude limit
in the B band (1); its median $M_B$ magnitude (2); the Schechter
characteristic magnitude in that redshift range (3); the difference
between the median and characteristic magnitudes (4); number of galaxies
(5); median redshift (6); measured correlation length and the slope of
the correlation function (7 and 8).

In the following sections we use -- at different redshifts
-- the characteristic magnitude of the Schechter
luminosity function in the 
$B$ band, $M_B^*$, as a reference value.  Normalizing our median absolute
magnitude values at each redshift to the corresponding value of
$M_B^*$ provides a way to take into account the mean brightening of
galaxies due to evolution, when comparing samples at different
epochs.  The values of $M_B^*$ at each redshift are those
estimated from these same data by Ilbert et al. (2005) in the AB system,
converted here into the VEGA scale.   

%
\begin{center}
 \begin{table*}
      \caption[]{Properties of the VVDS luminosity sub-samples in the
range $0.1<z<0.5$. 
}
   \begin{tabular}{|lcccrccc|}
            \hline
$M_B$ {\it range}  &  $M_B^{median}$ & $M_B^*$ & $M_B^{med}-M_B^*$ &  $N_{gal}$ & $z_{median}$ &   $r_0$ & $\gamma$ \\
            \hline
$< -16$ & -17.99 & -19.97 & 1.98 & 1330 & 0.332 & $2.55^{+0.54}_{-0.54}$ & $1.69^{+0.14}_{-0.17}$ \\
\hline
Volume limited: \\
\hline
$<-17.0$  & -18.30 & -19.97 & 1.67  & 1089 & 0.363 & $2.97^{+0.54}_{-0.46}$ & $1.72^{+0.16}_{-0.12}$ \\
$<-17.5$ & -18.53 & -19.97 & 1.44  & 883 & 0.373 & $3.17^{+0.62}_{-0.62}$ & $1.66^{+0.14}_{-0.11}$ \\
$<-18.0$ & -18.98 & -19.97 & 0.99 & 658 & 0.381 & $3.25^{+0.90}_{-0.92}$ & $1.72^{+0.22}_{-0.18}$ \\
$<-18.5$ &  -19.34 &  -19.97 & 0.63 &  475 & 0.388 & $3.47^{+0.90}_{-0.90}$ & $1.83^{+0.21}_{-0.20}$ \\
$<-19.0$ & -19.70 & -19.97 &  0.27 & 318 & 0.381 & $4.25^{+1.34}_{-1.54}$ & $1.70^{+0.27}_{-0.26}$ \\
$<-19.5$ & -20.00 & -19.97 & -0.03 &  201 & 0.391 & $3.65^{+2.26}_{-5.26}$ & $1.50^{+0.51}_{-0.40}$ \\
            \hline
   \end{tabular}
 \end{table*}
\end{center}
%
 \begin{table*}
      \caption[]{As Table 1, but for $0.5<z<1.2$ 
}   \begin{tabular}{|lcccrccc|}
            \hline
$M_B$ {\it range}  &  $M_B^{median}$ & $M_B^*$ & $M_B^{med}-M_B^*$ &  $N_{gal}$ & $z_{median}$ &   $r_0$ & $\gamma$ \\
            \hline
$<-17$ & -19.65 & -20.76 & 1.11 & 4283 & 0.808 & $2.75^{+0.27}_{-0.27}$ & $1.59^{+0.09}_{-0.07}$ \\
$<-18.5$ & -19.80 & -20.76 & 0.96 & 3736 & 0.856 & $2.89^{+0.27}_{-0.27}$ & $1.54^{+0.08}_{-0.07}$ \\
$<-19$ & -19.96 & -20.76 & 0.80 & 3272 & 0.882 & $2.95^{+0.33}_{-0.35}$ & $1.52^{+0.09}_{-0.08}$ \\
\hline
Volume limited: \\
\hline
$<-19.5$ & -20.23 & -20.76 & 0.53 & 2407 & 0.899 & $2.93^{+0.33}_{-0.35}$ & $1.59^{+0.12}_{-0.09}$ \\
$<-20$ & -20.58 & -20.76 & 0.18 & 1530 & 0.914 & $3.47^{+0.46}_{-0.43}$ & $1.84^{+0.14}_{-0.12}$ \\
$<-20.5$ & -20.92 & -20.76 & -0.16 & 865 & 0.913 & ${4.77}^{+0.61}_{-0.61}$ & ${2.00}^{+0.15}_{-0.12}$ \\
$<-21$ & -21.30 & -20.76 & -0.54 & 368 & 0.920 & $5.01^{+1.47}_{-1.65}$ & $2.38^{+0.36}_{-0.21}$ \\
            \hline
   \end{tabular}
 \end{table*}

\section{Estimating the real-space correlation function}

We summarize here the methods applied to derive the real-space
correlation function and its parameters, described
extensively in Pollo et al. (2005). The galaxy real-space correlation
length $r_0$ and slope $\gamma$ from the VVDS-F02 data are measured
from the projection of the redshift-space  correlation function
$\xi(r_p,\pi)$, estimated using the Landy \& Szalay (\cite{ls})
estimator,

\begin{eqnarray}
\xi(r_p,\pi) &\ = \ \frac{N_R(N_R-1)}{N_G(N_G-1)} \frac{GG(r_p,\pi)}{RR(r_p,\pi)} \\ \nonumber
       &\ - \frac{N_R-1}{N_G} \frac{GR(r_p,\pi)}{RR(r_p,\pi)} + 1, 
\label{lseq}
\end{eqnarray}
where $N_G$ and $N_R$ are the total numbers of objects in the galaxy 
sample and in a catalog of random points distributed within the same survey
volume and with the same redshift distribution and angular selection biases;
$GG(r_p,\pi)$ is the number of independent galaxy-galaxy pairs with
separation perpendicular to the line of sight between $\pi$ and $\pi+d\pi$ and separation parallel to the line of sight between $r_p$ and $r_p+dr_p$; $RR(r_p,\pi)$ is the number of independent
random-random pairs within the same interval of separations and $GR(r_p,\pi)$
represents the number of galaxy-random pairs. A total of $\sim 40 000$
random points has been used in each computation.

To derive the real-space correlation function and avoid the effect of
peculiar velocities which distort the redshift space statistics, we
integrate $\xi(r_p,\pi)$ along the line of sight (see \cite{davpeeb}),
computing the projected correlation function,
\begin{eqnarray}
w_p(r_p) & \equiv &  2 \int_0^\infty \xi(r_p,\pi) d\pi \\ \nonumber
 &  = & 2 \int_0^\infty
\xi\left[(r_p^2+y^2)^{1/2}\right] dy,
\label{wpdef}
\end{eqnarray}
where $\xi$ is the real space two-point correlation function
evaluated at the specific separation $r=\sqrt{r_p^2+y^2}$. In practice
the upper integration limit has to be chosen finite, to avoid adding
noise to the result. After a set of experiments we chose its
optimal value as $20$ $h^{-1}$ Mpc.  
If $\xi(r)$ is well described by a power law, $\xi(r) =
(r/r_0)^{-\gamma}$, 
the integral can be computed analytically, giving 
\begin{equation}
w_p(r_p) = r_p \left(\frac{r_0}{r_p}\right)^\gamma
\frac{\Gamma\left(\frac{1}{2}\right)\Gamma\left(\frac{\gamma-1}{2}\right)}
{\Gamma\left(\frac{\gamma}{2}\right)},
\label{wpmodel}
\end{equation}
where $\Gamma$ is Euler's Gamma function.  Fitting $w_p(r_p)$ for
separations $<10$ $h^{-1}$ Mpc using the procedure described in detail in
Pollo et al. (2005) provides a
best-fitting value of $r_0$ and $\gamma$ for each volume-limited
sub-sample.

The estimate of errors has been performed primarily using a bootstrap
resampling of the data.  However, our detailed error analysis in
Pollo et al. (2005), based on 50 VVDS mock surveys from the GalICS
simulations (\cite{blaizot}), indicates that bootstrap errors tend to
be an underestimate of the true ensemble errors.  The difficulty we
encounter is that luminosity-selected sub-samples of our mock surveys
do not show the same properties, in terms both of total numbers and
scaling of the intrinsic clustering with luminosity, as the real VVDS.
We have no guarantee, therefore, that for this specific application
the variance among the mock samples represent a realistic
estimate of the errors in the real data.  For this reason, we have
decided here to estimate error bars on $w_p(r_p)$ using the bootstrap
technique, but correct these empirically to include the contribution
of cosmic variance.  An average value for this correction has been
estimated from the direct comparison of the errors computed in both
ways for 50 whole mock samples (\cite{meneux}, \cite{techcorr}).  The
overall mean effect is to increase the size of the error bars on $w_p$
by a factor of $\sim 2$ for the low-redshift samples, and by a factor
of $\sim 1.3$ for the high-redshift samples.  We applied this
correction to all our bootstrap estimates of $w_p(r_p)$.
We have checked that our conclusions are robust
to the details of this correction: even doubling the error bars in
both redshift ranges, the trends in the values of correlation length
and slope that we find at $z\sim 1$ do not change and remain
significant.

\section{Results}

\begin{figure*} \centering
   \includegraphics[width=14cm]{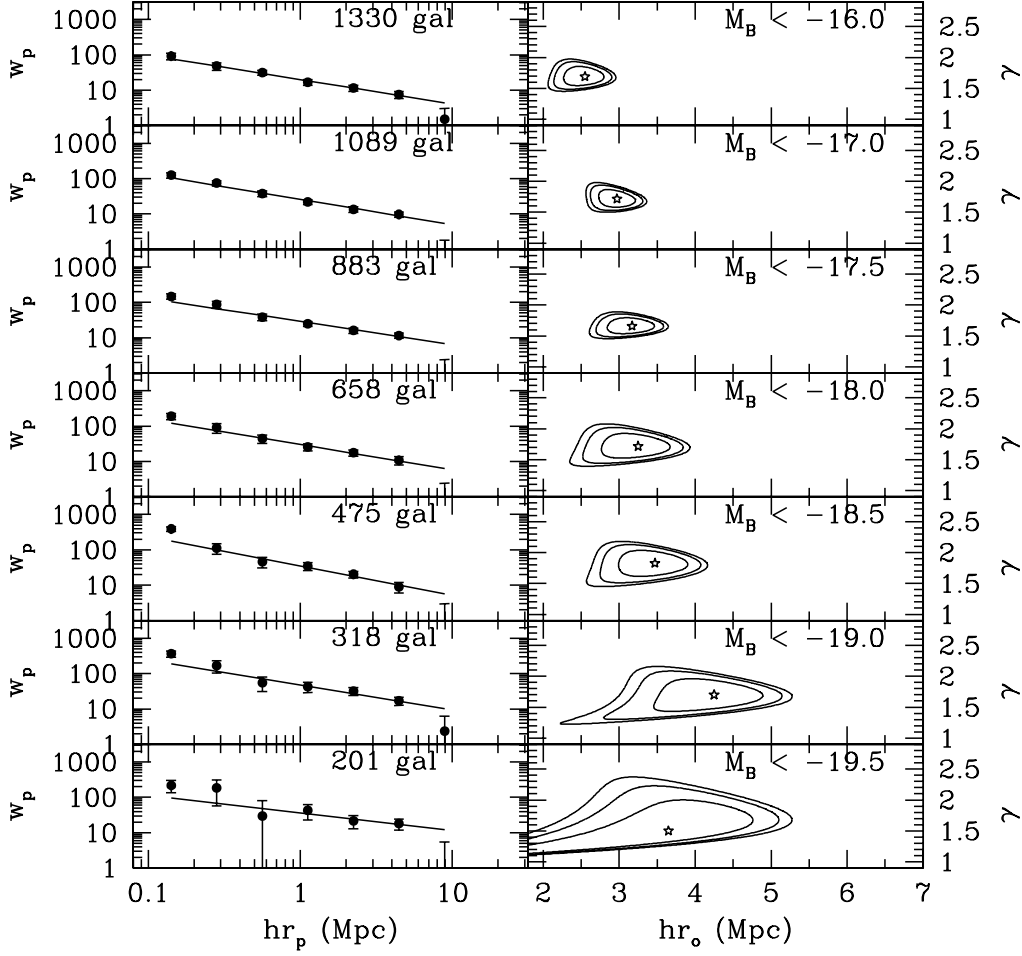} \caption{Projected
   correlation functions $w_p(r_p)$ and the associated $r_0$ and
   $\gamma$ error countours for the seven volume limited catalogs at
   $z \le 0.5$.}  \label{lowz_cont} \end{figure*}

\begin{figure*} \centering
   \includegraphics[width=14cm]{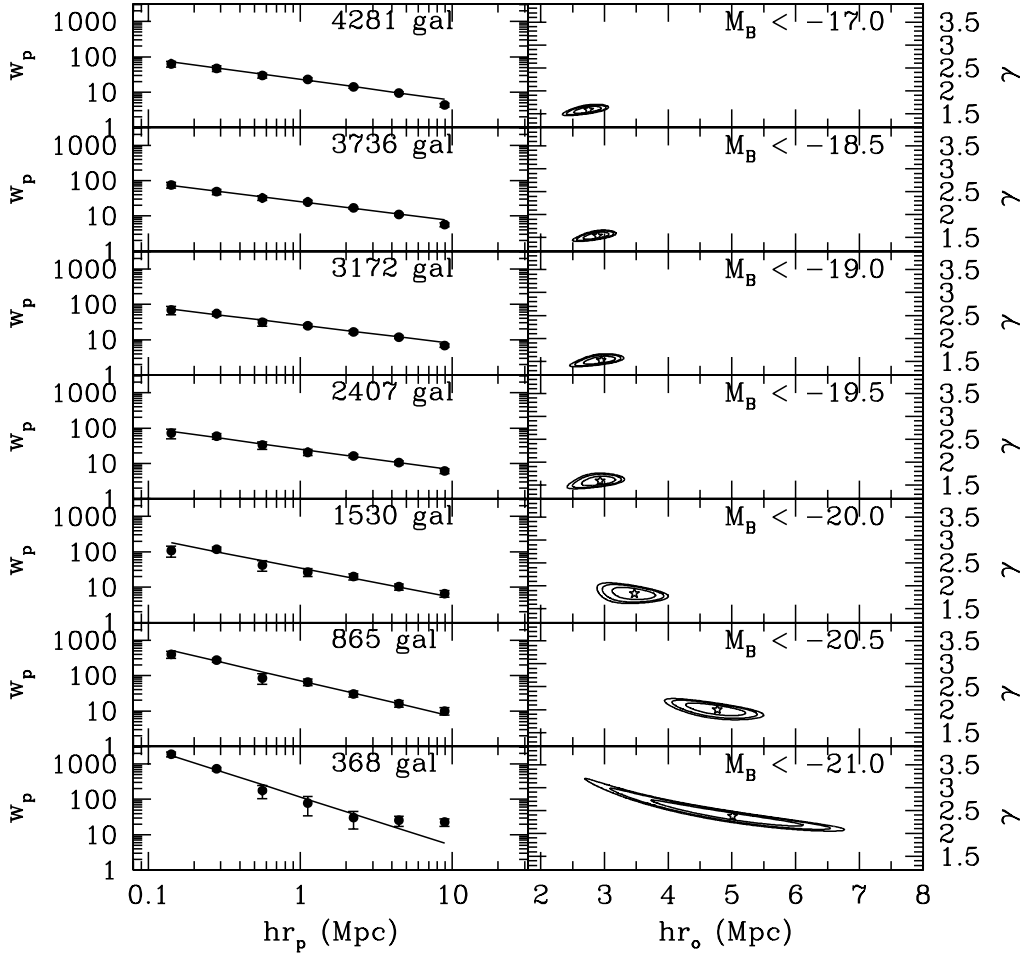}
   \caption{Projected correlation functions $w_p(r_p)$ and the
   associated $r_0$ and $\gamma$ error countours for the seven
high-redshift ($0.5 < z \le 1.2$) samples.}  \label{highz_cont} \end{figure*}

Figures ~\ref{lowz_cont} and~\ref{highz_cont} show the results of the
power-law fits of the projected correlation functions $w_p(r_p)$ and the
corresponding $r_0$ and $\gamma$ error contours, for the low- and
high-redshift samples.  The fitting has been performed taking into
account the full covariance matrix of the data, as described in
Pollo et al. (2005). 

\subsection{The correlation length $r_0$}

A comparison of Figures~\ref{lowz_cont} and \ref{highz_cont}
shows qualitatively that for the high-redshift samples 
both the correlation length 
and the slope of the correlation function change with the
sample luminosity.   Note that we do not perform here any 
analysis of the detailed shape of $w_p(r_p)$, its evolution and its
implications for halo occupation models, but limit ourselves to the
simple and robust fit of $w_p(r_p)$ with a single power law.

\begin{figure*}
   \centering
     \includegraphics[width=8.5cm]{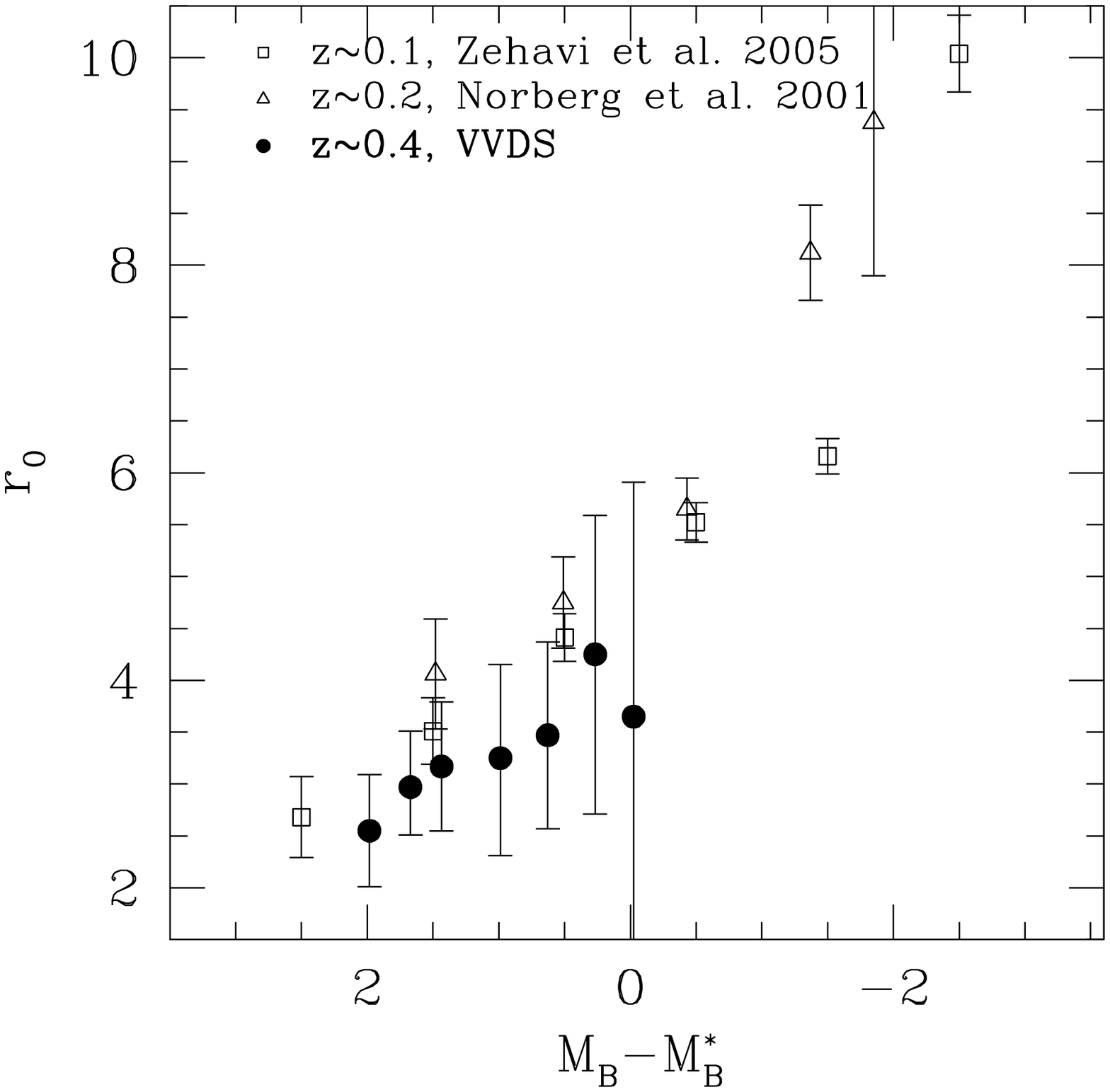}
     \includegraphics[width=8.5cm]{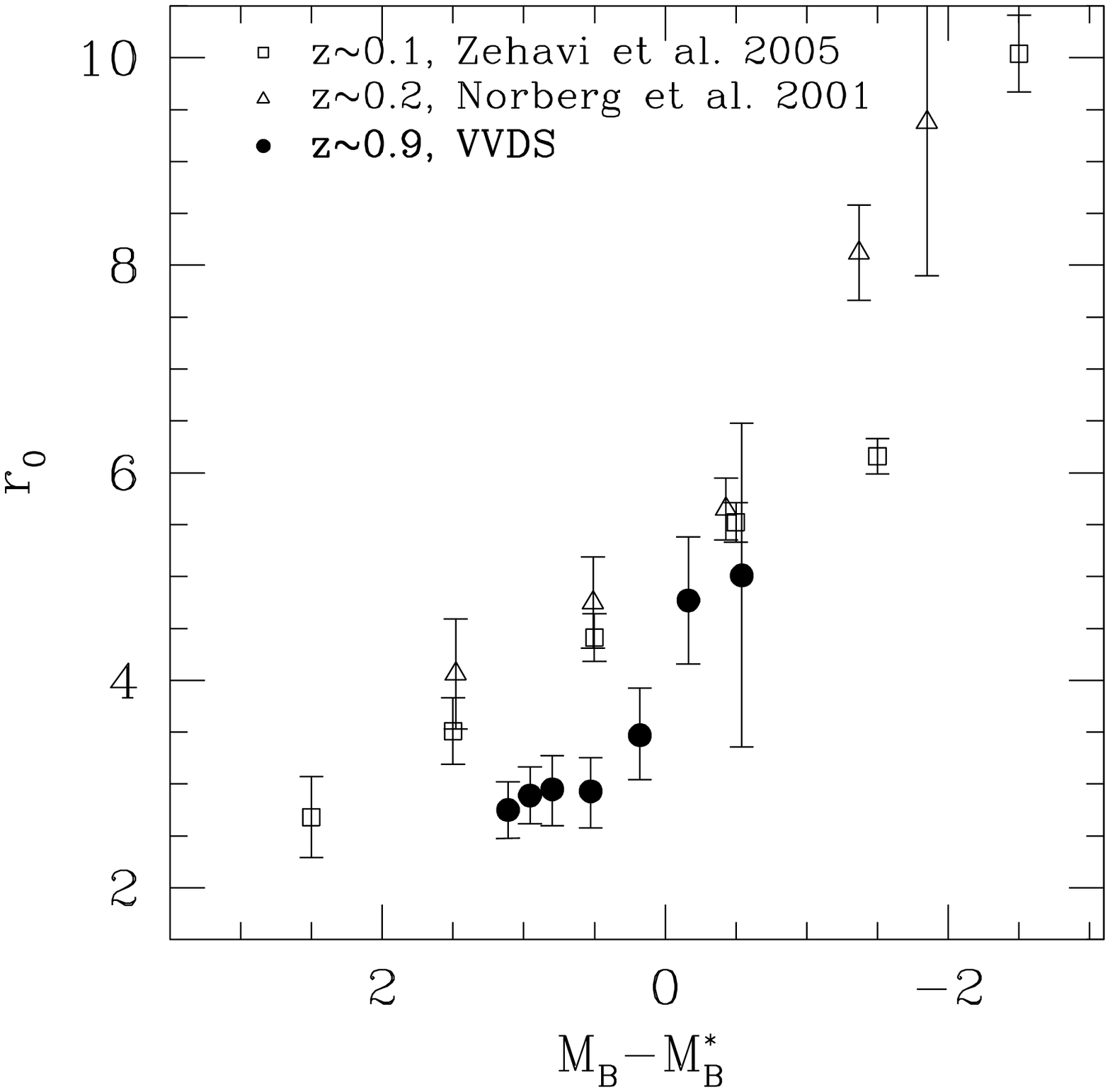}
      \caption{
Left panel: dependence of the clustering length $r_0$ on the median
luminosity of galaxies in the local Universe from
2dFGRS and SDSS compared to the VVDS measurements at $z \sim 0.4$.  Right panel: the same local reference values from
2dFGRS and SDSS compared to the VVDS measurements at
$\left<z\right>\sim 0.9$. } 
\label{comp} \end{figure*}

In the left panel of Figure~\ref{comp} we compare our measurements of
$r_0$ from the low-redshift VVDS samples to the
similar measurements from the 2dFGRS
(\cite{norberg}) and the SDSS (\cite{zehavi}).  A one-to-one
comparison is not possible, since the large size of local surveys
allowed for the measurements in a series of disjoint volume-limited
surveys with small magnitude intervals, i.e. $L1 \le L \le L2$,
with $L1$ and $L2$ being their limiting luminosities.
In our
case we are forced to use integral measurements, i.e. samples with $L
\le L2$.
Still,
different sub-samples are dominated by galaxies with a specific
characteristic luminosity that we characterize by computing the
median absolute magnitude within each sample (reported in Tables 1 and
2).  In addition, given the smaller volume of our low-redshift
samples, it is clear that rare luminous objects will be
under-represented.  Finally, our low-redshift samples extend to
$z=0.5$ and thus 
may be regarded as a subsequent redshift bin after 2dFGRS and SDSS.  
Considering these intrinsic limitations ,
the overall trend is consistent with
that observed by 2dFGRS and SDSS, although our measurements
are systematically lower.

With this consistency in mind we analyze our
VVDS high-redshift ($z\simeq0.9$) samples and compare it to the local
values from 2dFGRS and SDSS. 
This is done in the right panel of Figure~\ref{comp}.  
Galaxies fainter than $M_B^*$ at high redshift are significantly less
clustered than their counterparts in the present-day Universe, with
$r_0=2.75\pm0.27$ $h^{-1}$ Mpc. At the same time, the clustering
strength of galaxies brighter than $M_B^*$ is comparable to that
observed locally with a correlation length up to $r_0=4.77\pm0.61$
$h^{-1}$ Mpc. We therefore observe that at redshift $z\simeq0.9$, as
luminosity increases above $L^*$, the clustering length 
suddenly rises to values comparable to those observed locally for
galaxies with similar $M_B-M^*$.   

\subsection{The correlation function slope $\gamma$}

\begin{figure*} \centering
   \includegraphics[width=8.5cm]{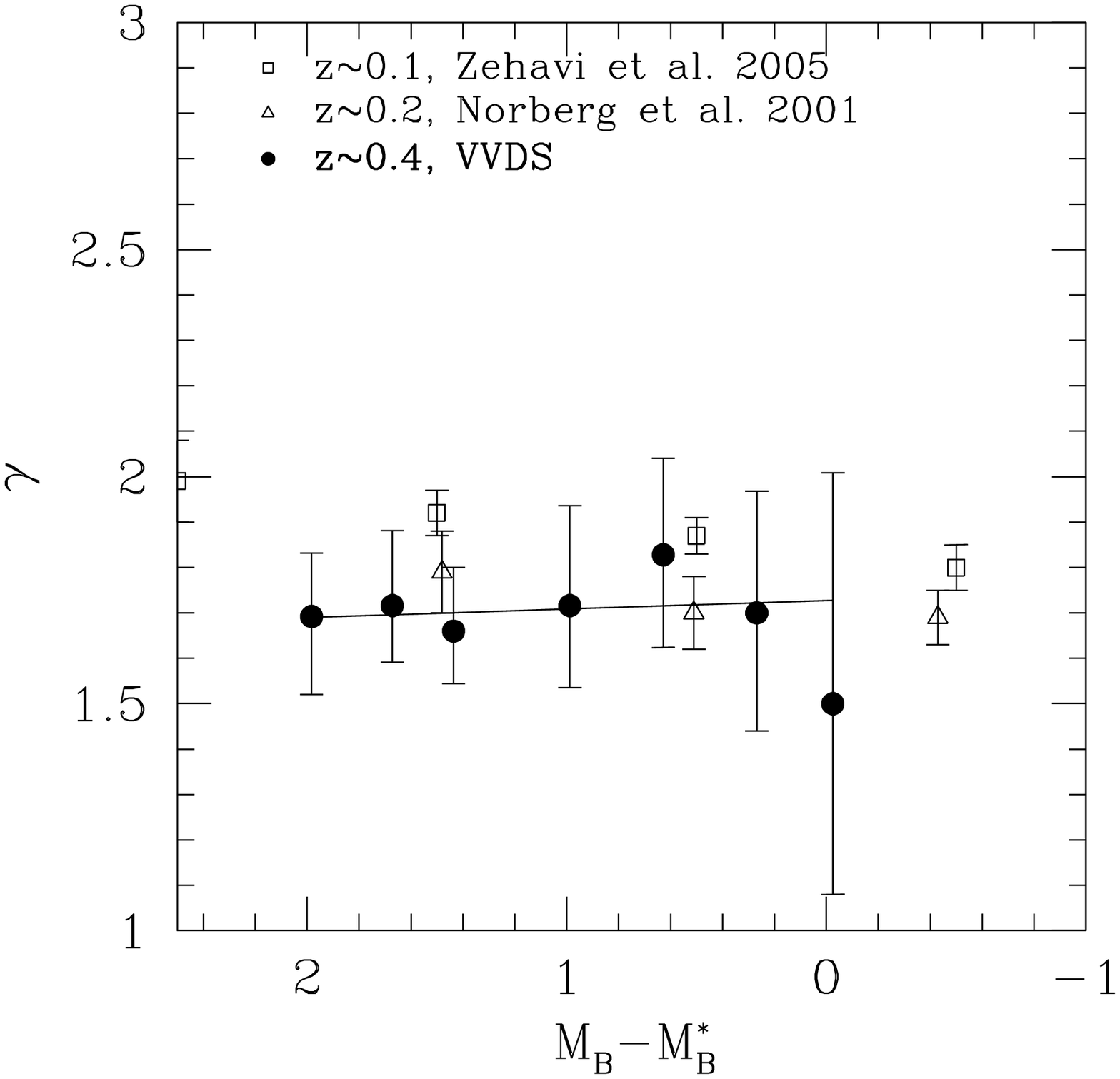}
   \includegraphics[width=8.5cm]{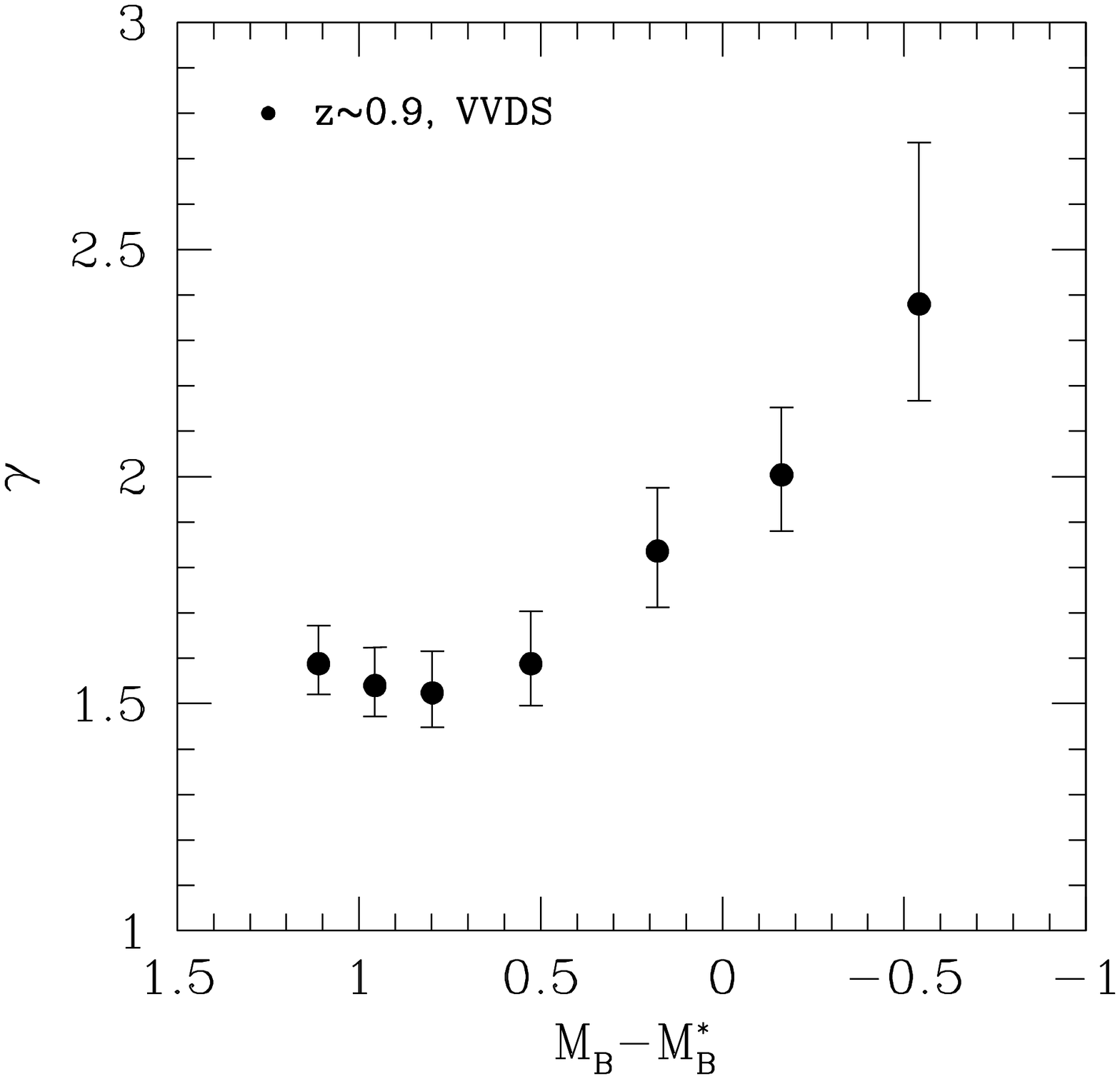} \caption{Measured
values of the slope of $\xi(r)$, $\gamma$, as a function of the
median luminosity of galaxies at low redshift (left panel) and at $z \sim 0.9$
   (right panel). While locally and up to $z \sim 0.5$ $\gamma$ 
remains practically
   constant (with the best rms fit $\gamma = 1.73 - 0.02(M_B-M_B^*)$,
marked as a solid line), 
in the distant Universe it shows a clear scaling with
   luminosity for galaxies brighter than $M_B^*+0.5$: the spatial
correlation function becomes steeper for galaxies of
increasing luminosity.} \label{gamma}
   \end{figure*}

As we show in Figure~\ref{gamma}, the dependence of the correlation
function slope $\gamma$ on the galaxy intrinsic luminosity in the
high-redshift sample differs strongly from the local
measurements. Locally, $\gamma$ has a remarkably constant value, with
$\gamma\simeq 1.7$ measured both by the much larger local surveys 
and - in a very consistent way - by VVDS in our closer redhift bin 
$z \sim 0.4$. 
Conversely, at high $z$, $\xi(r)$ clearly steepens with 
increasing luminosity for all sub-samples brighter than $M_B\simeq M_B^*+0.5$.
We find that for $M_B - M_B^* > 0.5$, the slope of the correlation
function is consistent with $\gamma=1.55\pm0.08$, while for 
$M_B - M_B^* < 0$ the slope goes up to $\gamma=2.38^{+0.36}_{-0.21}$.

\subsection{The relative bias of different luminosity classes}

   \begin{figure*}
   \centering
   \includegraphics[width=8.5cm]{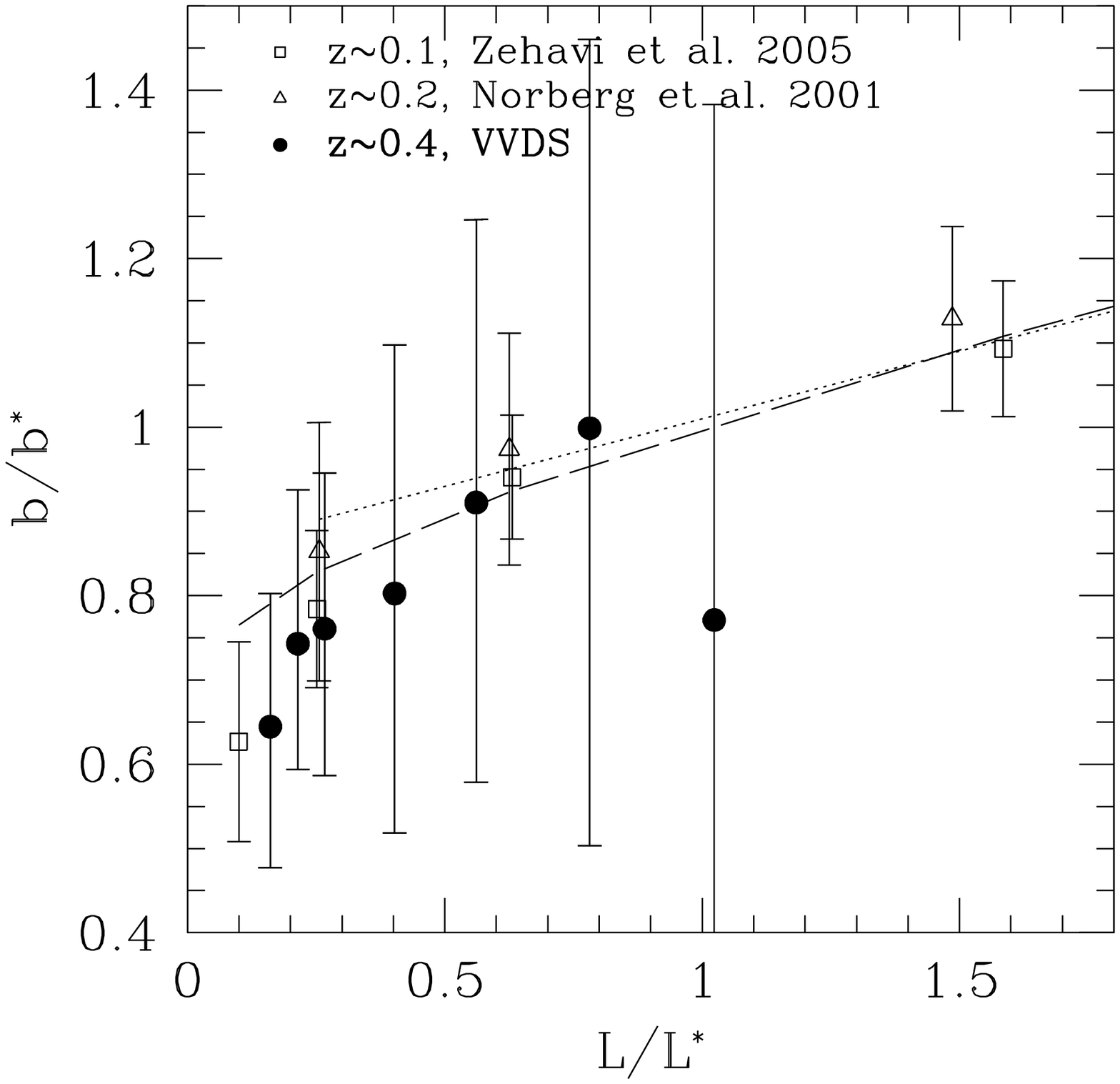}
   \includegraphics[width=8.5cm]{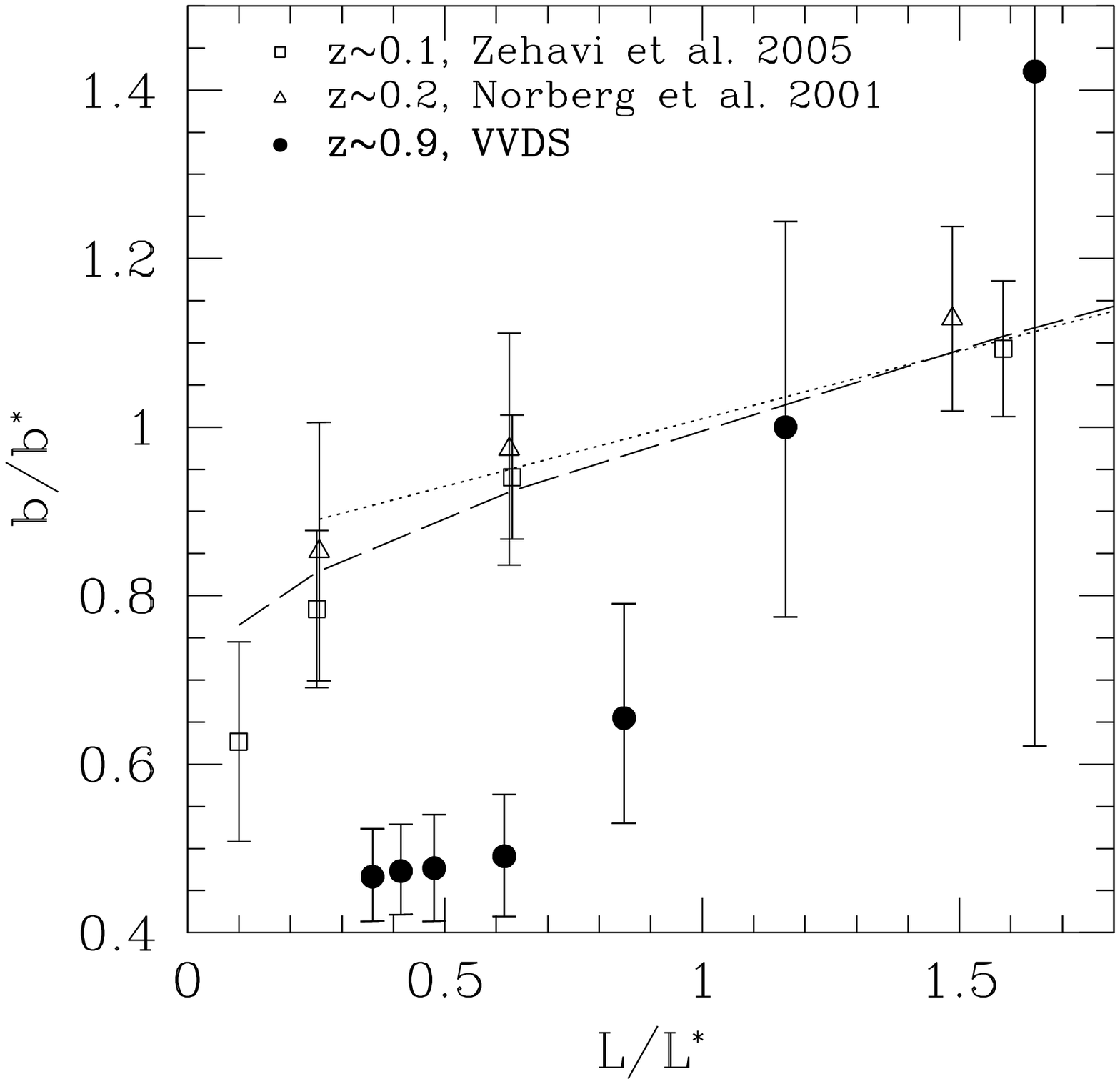} \caption{The
   variation of the relative bias in the VVDS sub-samples (filled
   circles) as a function of luminosity using the clustering of $L^*$
   galaxies as a reference point, compared to the 2dFGRS (open triangles)
   and SDSS (open squares) local results. The dashed and dotted lines
   show, respectively, the best fits made for the SDDS and 2dFGRS
   measurements. } \label{bias} \end{figure*}

To interpret our results and to
compare them to local surveys,  we compute the relative bias parameter,
$b/b^*$, which gives the amplitude of the correlation function
relative to that of $L^*$ galaxies. 
Consistently with the 2dFGRS analysis
(e.g. \cite{norberg02}) we define the relative bias of the generic $i$-th
sample with a given median luminosity $L$, with respect to that
corresponding to $L^*$, as

\begin{equation}
\frac{b_i}{b^*} = \sqrt{\frac{(r_0^i)^{\gamma_i}}{(r_0^*)^{\gamma^*}}r^{\gamma^*-\gamma^i} }\,\,\,\,  ,
\label{rbias}
\end{equation}
and estimate it at the fixed $r=1$ $h^{-1}$ Mpc scale (see also Meneux
et al. 2006 for a slightly different definition).

To apply this formula to the low-redshift samples, we need to estimate
the values of $r_0$ and $\gamma$ for $M_B^*$ galaxies, that we obtain
by a linear fit to the observed VVDS data in the left panel of Figure~\ref{comp} 
(avoiding the very uncertain value measured at $M_B^*$).  
We then plot the corresponding values of $b/b^*$ for our samples
in Figure \ref{bias}.
In this figure we also plot the 2dFGRS and SDSS data, together with the
analytic fitting relations provided for them, $b/b^* = 0.85 + 0.16
L/L^*$ for the 2dFGRS (\cite{norberg02}) and $b/b^* = 0.85 + 0.15 L/L^* - 0.04 (M-M^*)$
for the SDSS (\cite{tegmark}).  Given the error bars, we can say that the
low-redshift VVDS results are in good agreement with both the SDSS
and the 2dFGRS fits.  

In the high-redshift samples the situation is clearly different. As we
can see from the right panel of Fig.~\ref{bias}, 
the value of $b/b^*$ 
rises steeply from low values $b/b* \simeq 0.5$ at low luminosities to values that,
statistically, are consistent with those of the local samples
for $L>L^*$ galaxies, $b/b* \simeq 1 - 1.4$.   At the same time 
the difference on relative bias 
between
of galaxies fainter and brighter than $L^*$
becomes very large.  This 
appears to be an indication that going
back in time the bias contrast of luminous galaxies to the rest of the
population becomes stronger and is consistent with the fact that
fainter galaxies are found to be significantly less biased tracers
of the mass than the L* galaxies even at this relatively high redshift 
(\cite{marinoni}).

\section{Summary and discussion}

The projected correlation functions that we have measured from our set
of volume limited sub-samples of the VVDS are in general
fairly well fitted by a single power-law in the range $0.1 \le
r/h^{-1} {\rm Mpc} \le 10$, both for the low-redshift and high-redshift
samples.  This allows us to use variations in $r_0$ and $\gamma$ to
characterize the global dependence of clustering on luminosity at
high redshift and compare it to similar low-redshift results.  The
observed behaviour has strong implications for HOD models,
as it directly impacts on any recipe for populating dark matter halos
at high redshift. Deviations
from the power-law shape, although extremely interesting for
constraining  HOD models (e.g. \cite{zehavi}) are not analyzed
in this paper and will be the subject of future work. 

We 
observe that for median redshifts $z\simeq 0.9$ ($0.5<z<1.2$) 
the clustering length has a low, nearly constant value
$r_0\simeq 2.9$ $h^{-1}$ Mpc for luminosities $M_B<M_B^*$, 
rising suddenly for $M_B>M_B^*$ and approaching values 
$r_0\simeq 5$ $h^{-1}$ Mpc similar to those of
local galaxies with comparable luminosity (relatively to the
characteristic value $M_B^*$, \cite{norberg}). 
This behaviour is consistent with general predictions by
hierarchical models of galaxy formation (\cite{benson}), where
luminous galaxies are more confined to the peaks of the
large-scale density field, going back in redshift, simply due to the
higher bias of the parent halos. 

Another important result of this work is the
clear detection of a systematic steepening of the high-redshift
correlation function for absolute magnitudes brighter than $\sim
M_B^*+0.5$ (Figure
\ref{gamma}). This kind of behaviour is in general not seen either in our 
closer $z \sim 0.4$
sample or in 
the large local surveys, although Zehavi et al. (2005) do detect an
increase of $\gamma$ in their most luminous volume-selected sample. 
A similar trend with increasing (UV) luminosity has been recently
observed for a population of Ly-break galaxies at $z \ge 4$ in the
Subaru Deep Field (\cite{kashikawa}).  These authors are able to reproduce their
observed relationship with a HOD model in which they introduce
multiple LBGs into massive dark matter haloes. This 
amplifies the clustering strength at small scales, steepening the
correlation function.  A similar interpretation could be applied to our data.
Finally, the observed relative bias of galaxies at
high redshift provides evidence for a clear difference in the
clustering properties of galaxies fainter or brighter than the
characteristic luminosity:
sub-samples with $M_B \lesssim M_B^*$ behave in a way that is very
similar to local samples while the relative bias of samples with $M_B \gtrsim M_B^*$ remains
significantly lower.   

Results presented in this paper show that there is a significant redshift evolution of the luminosity dependence of
both the normalization and slope parameter of the galaxy 
correlation function. 
This specific observation can provide
an important test of galaxy formation models,
constraining in particular the multiplicity of luminous galaxies
within massive halos at $z=1$. 

NOTE ADDED IN PROOF: In a parallel paper \cite{coil05} perform 
a similar measurement at $z \sim 1$ using the DEEP-2 survey. Although
they explore a narrower range in median luminosities, they also detect
a first hint of the steepening of $w_p(r_p)$ above $M_B^*$ and a rise of
the correlation length with luminosity. Considering cosmic variance 
and the different selection function (unlike the VVDS, DEEP-2 is not 
a purely magnitude-limited survey), the overall results from these 
two data sets are thus in good agreement.

\begin{acknowledgements}
We thank Peder Norberg for very useful comments and suggestions.  \\
This research has been
developed within the framework of the VVDS consortium.\\ This work has
been partially supported by the CNRS-INSU and its Programme National
de Cosmologie (France), and by Italian Ministry (MIUR) grants
COFIN2000 (MM02037133) and COFIN2003 (num.2003020150).\\ The VLT-VIMOS
observations have been carried out on guaranteed time (GTO) allocated
by the European Southern Observatory (ESO) to the VIRMOS consortium,
under a contractual agreement between the Centre National de la
Recherche Scientifique of France, heading a consortium of French and
Italian institutes, and ESO, to design, manufacture and test the VIMOS
instrument.
\end{acknowledgements}

\end{document}